\documentclass[twocolumn,twoside,preprintnumbers,amsmath,amssymb,pacs,aps]{revtex4}
\usepackage{epsfig}
\usepackage{graphicx}
\usepackage{dcolumn}
\usepackage{bm}

\usepackage{fancyhdr}

\usepackage{pslatex}

\begin{document}

\title{{\Large  Langevin dynamics of the deconfinement transition 
for pure gauge theory}}

\author{Ana J\'ulia {\sc Mizher}$^1$\footnote{anajulia@if.ufrj.br}, 
Eduardo S. {\sc Fraga}$^1$\footnote{fraga@if.ufrj.br},
Gast\~ao {\sc Krein}$^2$\footnote{gkrein@ift.unesp.br} }
\affiliation{$^1$Instituto de F\'\i sica, 
Universidade Federal do Rio de Janeiro, Caixa Postal 68528, 
21941-972 Rio de Janeiro, RJ , Brazil\\
$^2$Instituto de F\'\i sica Te\'orica, Universidade Estadual Paulista, 
Rua Pamplona 145, S\~ao Paulo, SP 01405-900, Brazil}

\received{on 31 March, 2006}

\begin{abstract}

We investigate the effects of dissipation in the deconfinement 
transition for pure $SU(2)$ and $SU(3)$ gauge theories. Using an 
effective theory for the order parameter, we study its Langevin 
evolution numerically. Noise effects are included for the case of 
$SU(2)$. We find that both dissipation and noise have dramatic effects 
on the spinodal decomposition of the order parameter and delay 
considerably its thermalization. For $SU(3)$ the effects of dissipation 
are even larger than for $SU(2)$.

\vspace{0.5cm}



\end{abstract}

\maketitle

\thispagestyle{fancy}
\setcounter{page}{0}


\section{Introduction}

Recent results from lattice QCD \cite{laermann}, corroborated by 
experimental data from BNL-RHIC \cite{bnl}, indicate that 
strongly interacting matter under extreme conditions of temperature and 
pressure undergoes a phase transition to a deconfined plasma. Such 
extreme conditions are believed to have happened in the early universe, 
and might also be found in the core of neutron stars \cite{G_b}.

The process of phase conversion during the deconfinement transition can 
occur in different ways. For a pure gauge $SU(N)$ theory, the 
trace of the Polyakov loop provides a well-defined order parameter 
\cite{polyakov,thooft,pisarski}, and one can construct an effective 
Landau-Ginzburg field theory based on this quantity 
\cite{pisarski2,ogilvie}. The effective potential for 
$T<<T_c$ has only one minimum, at zero, where the whole system is localized. 
With the increase of the temperature new minima appear ($N$ minima 
for $Z(N)$, the center of $SU(N)$). At the critical temperature, 
$T_c$, all the minima are degenerate, and above $T_c$ the new  
minima become the true vacuum states of the theory, so that the system 
starts to decay. In the case of $SU(3)$, whithin a range of temperatures 
close to $T_c$ there is a small barrier, and the process of phase 
conversion will be guided by bubble nucleation. For larger $T$, the 
barrier disappears and the system explodes in the process of spinodal 
decomposition. For $SU(2)$, the transition is second-order, and there 
is never a barrier to overcome. 

In this paper, we consider pure $SU(2)$ and $SU(3)$ gauge theories, 
without dynamical quarks, that are rapidly driven to very high 
temperatures, well above $T_c$, and decay to the deconfined phase via 
spinodal decomposition. We are particularly interested in the effect of 
noise and dissipation on the time scales involved in this ``decay 
process''. In what follows, we adopt an effective model proposed in 
Ref. \cite{ogilvie} for the order parameter and the 
effective potential. Numerical calculations for the evolution of the 
order parameter are performed on a lattice, using a local Langevin 
equation. 

The paper is organized as follows. Section II briefly describes 
the effective model for the order parameter. In Section III, we 
consider the Langevin evolution, discussing how to fix the 
dissipation coefficient from lattice simulations, and present our 
results for $SU(2)$ and $SU(3)$. Section IV contains our final remarks.


\section{The effective model}

The model proposed in \cite{ogilvie} intends to provide a better 
representation of lattice results for the gluon plasma equation 
of state as compared to the usual bag model. 
It is obtained combining a few phenomenological inputs with $Z(N)$ symmetry 
and some known features of the perturbative equation of state. In particular, 
in a temperature range going from the deconfinement temperature $T_d$ to $5T_d$ 
the model gives reasonable results and exhibits a thermodynamic behaviour 
that is coherent with data obtained from lattice simulations.

In this approach, thermodynamic properties are determined by functions
of the Polyakov loop, defined in Euclidean finite temperature gauge 
theories as \cite{polyakov}: 
\begin{equation}
P(\vec x )= {\cal T}exp\left[ig\int_0^{1/T}d\tau ~A_0(\vec x, \tau)
\right] \; ,
\end{equation}
where ${\cal T}$ denotes Euclidean time ordering, $g$ is the 
gauge coupling constant and $A_0$ is the time component of the vector 
potential. We work with $SU(2)$ and $SU(3)$, 
representing the color degrees of freedom. Consequently, we have a $Z(N)$ 
symmetry for the case of pure gauge theories that is spontaneously 
broken. It would be explicitly broken in the presence of quarks.

Working in the imaginary time framework, we have bosonic fields being 
periodic and fermionic fields being antiperiodic in the 
imaginary time $\tau$:
\begin{equation}
A_\mu (\vec x, \beta)=+ A_\mu (\vec x,0)\ \ , \ q(\vec x, \beta) = 
-q(\vec x , 0) \; .
\end{equation}
Any gauge transformation periodic in $\tau$ respects these boundary 
conditions. However, as demonstrated by 't Hooft \cite{thooft}, one can
consider more general gauge transformations which are only periodic up
to the center of the group: 
$\Omega (\vec x, \beta)= \Omega_c\ \ , \ \ \Omega(\vec x,0)=1$.

Color adjoint fields are invariant under these transformations, while 
those in the fundamental representation are not:
\begin{equation}
A^\Omega(\vec x, \beta)=\Omega^\dag_c A_\mu (\vec x, \beta)\Omega_c=
A_\mu(\vec x, \beta) = + A_\mu(\vec x, 0) \; ,
\end{equation}
\begin{equation}
q^\Omega (\vec x, \beta) = \Omega ^\dag_c q(\vec x, \beta ) 
\neq -q(\vec x, 0) \; .
\end{equation}
Consequently, pure gauge theories have a global $Z(N)$ symmetry, which 
is spoiled by the addition of dynamical quarks.

Thus the action is invariant under $Z(N)$ transformations, but $\langle 
{\rm Tr}_F \, P(\vec x)\rangle$ is not. Symmetry requires 
$z \, \langle {\rm Tr}_F \, P(\vec x)\rangle = \langle {\rm Tr}_F \, 
P(\vec x)\rangle$, 
which implies $\langle {\rm Tr}_F \, P(\vec x)\rangle=0$. 
When the phase transition occurs, the $Z(N)$ symmetry is spontaneously 
broken and  $\langle {\rm Tr}_F \, P(\vec x)\rangle$ assumes a non-vanishing 
value. So, one can use it as an order parameter for the transition and 
it is possible to write an effective theory for its dynamics. 

The efective theory of Ref. \cite{ogilvie} is based on a mean 
field treatment in which the Polyakov loops are constant throughout the space 
and the free energy is a function of its eigenvalues. A perturbative 
calculation of the free energy of gluons as a function of the Polyakov loop 
eigenvalues yields
\begin{equation}
f = -\frac{1}{\beta}\sum^{N}_{j,k=1}2\left(1-\frac{1}{N}\delta_{jk}\right)
\int\frac{d^3 k}{(2\pi)^3}\sum_{n=1}^{\infty}\frac{1}{n} e^{- \beta m  w_k +
in \Delta\theta_{jk}} \; ,
\end{equation}
where $\theta$ is defined through the eigenvalues of the Polyakov loop, 
$P_{jk} = \exp(i\theta_j) \; \delta_{jk}$, and 
$\Delta\theta_{jk}\equiv \theta_j-\theta_k$, 
which reduces to the usual blackbody formula in the case $A_0=0$.

This expression for the free energy of gluons propagating in the 
background of Polyakov loops predicts a gas of gluons that is always 
in the deconfined phase, with no indication that higher-order 
corrections will modify this result. This can be modified by the 
introduction of a mass scale into $f$. This mass scale will 
determine the deconfinement temperature $T_d$ and is introduced in a 
phenomenological way. One ends up with the same expression of the 
perturbative calculation, but now $w_k=\sqrt{k^2+M^2}$. Parametrizing 
the Polyakov loop and representing the diagonal matrix as 
$diag[\exp(i\phi_{N/2},...,i\phi_1,-i\phi_1,...,-i\phi_{N/2}]$, 
it is possible to extract an effective potential as a function of 
$\phi$. For $SU(2)$ one has only $\phi_1=\phi$ and $\phi_{-1}=-\phi$, and 
the effective potential can be written as
\begin{eqnarray}
V &=& -\frac{\pi^2T^3}{15} + \frac{4T^3}{3\pi^2}\phi^2(\phi-\pi)^2 \nonumber \\
&+& \frac{M^2T}{4} + \frac{M^2T}{\pi^2}\phi(\phi-\pi) \; .
\end{eqnarray}
Notice that there is a symmetry $\phi\leftrightarrow \pi-\phi$  
associated with the $Z(2)$ invariance. It is convenient to write this 
equation in terms of a new variable $\psi=1-\phi\pi/2 $ to make this 
symmetry more evident. One obtains:
\begin{eqnarray}
V &=& -\frac{\pi^2T^3}{15}+\frac{T^3\pi^2}{12}(1-\psi^2)^2\nonumber \\
&+& \frac{M^2T}{4} -\frac{M^2T}{4}(1-\psi^2) \; ,
\end{eqnarray}
where $\psi=0$ represents confinement. 
It is interesting to connect the $\psi$ used here and the trace of the 
Polyakov loop, used in Ref. \cite{pisarski2} as the order parameter. 
For the diagonalyzed matrix 
we have the trace, according to our parametrization, as 
$e^{i\phi}+e^{-i\phi}$,
namely ${\rm Tr} \, L = 2 \cos(\phi)$, or, as defined above, 
${\rm Tr} \, L = 2 \cos(\pi (1-\psi)/2)$. So, when $\psi =0$ we have
${\rm Tr} \, L = 0$, which represents confinement, and when 
$\psi\rightarrow 1$, then ${\rm Tr} \, L\rightarrow 1$, representing 
the deconfined state.

The phase transition in this case is second order, as expected \cite{yaffe}. 
The value of $M$ can be determined from the 
deconfinement temperature through the relation
$T_d=(3/2)^{1/2}M/\pi \approx 0.38985M$, 
so that it is possible to extract the deconfining temperature from 
the lattice and then fix $M$. The minimum of the potential occurs 
for
\begin{equation}
\psi_0=\sqrt{1 - \frac{3M^2}{2T^2\pi^2}} \; .
\end{equation}

For $SU(3)$ there are three eigenvalues: $\phi_1=\phi, \ 0$ 
and $\phi_{-1}=-\phi$. The potential assumes the form:
\begin{eqnarray}
V &=& - T^3\frac{8\pi^2}{45} + \frac{T^3}{6\pi^2}[8\phi^2(\phi-\pi)^2+
\phi^2(\phi-2\pi)^2] \nonumber\\
&+& \frac{2TM^2}{3} +\frac{TM^2}{2\pi^2}[2\phi(\phi-\pi)+\phi(\phi-2\pi)]\; .
\end{eqnarray}
Again, it is useful to rewrite the potential in terms of a new variable 
$\psi=2\pi/3-\phi$, so that one obtains
\begin{eqnarray}
V &=& \frac{8\pi^2}{405}T^3+\left(\frac{3}{2\pi^2}TM^2-\frac{2}
{3}T^3\right)\psi^2 \nonumber\\ 
&-&\frac{2}{3\pi}T^3\psi^3+\frac{3}{2\pi^2}T^3\psi^4 \; .
\end{eqnarray}
Now, $M$ and $T_d$ are related as follows:
\begin{equation}
T_d=\frac{9}{20\pi}\sqrt{10}M\approx 0.45296 \, M \; ,
\end{equation}
and the minimum is at
\begin{equation}
\psi_0 =\frac{\pi T+3\sqrt{T^2\pi^2-2M^2}}{6T} \; .
\end{equation}
In this case, ${\rm Tr} \, L = e^{i\phi} + 1 + e^{-i\phi}$ and the 
connection 
with $\psi$ becomes ${\rm Tr} \, L = \frac{2}{3} \cos\left(\frac{2\pi}{3}
-\psi\right) +\frac{1}{3}$.
%
%
%
\vspace{0.7cm}

\begin{figure}[htbp]
\includegraphics[width=7.75cm]{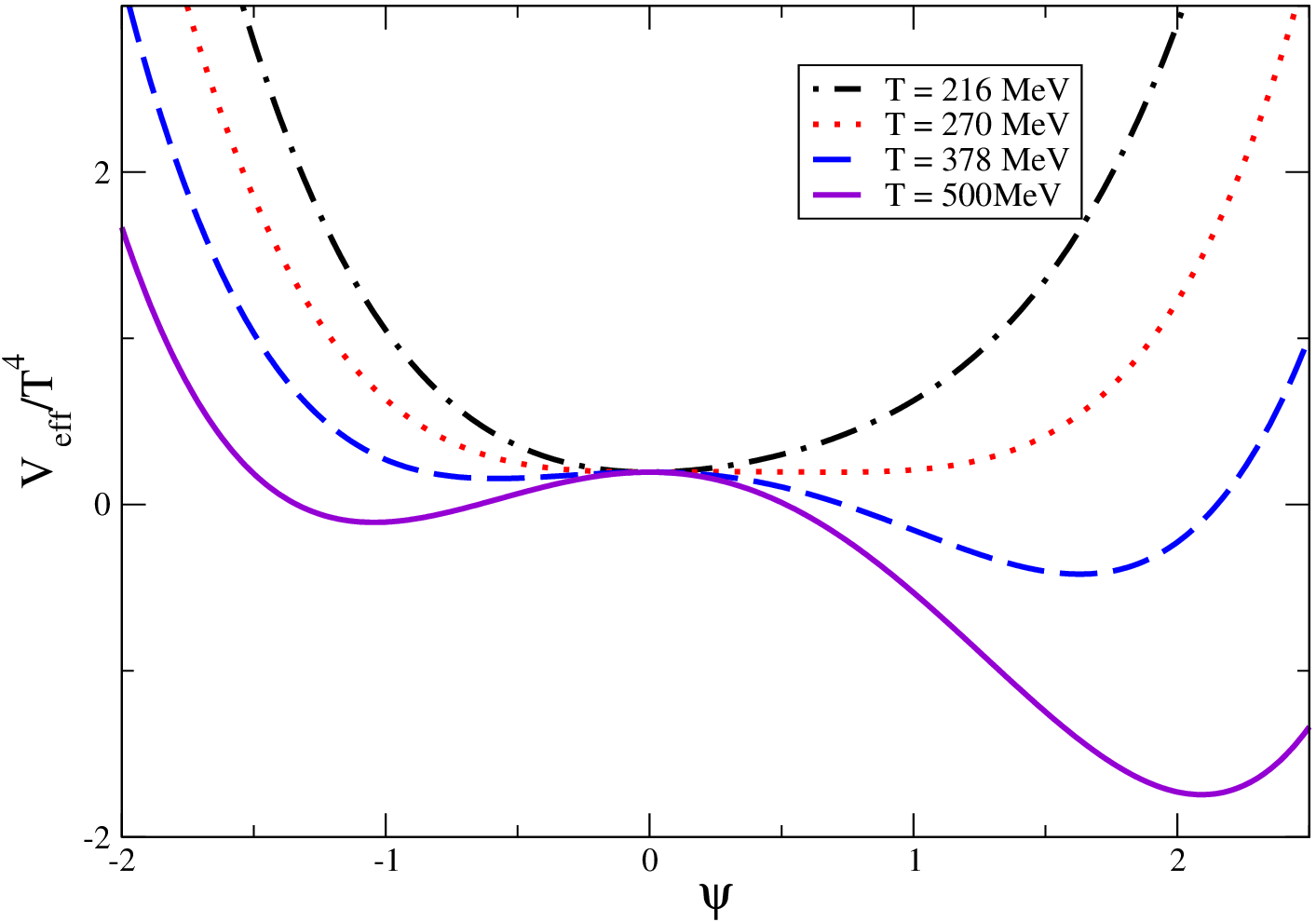}
\vspace{0.10cm}
\caption{Effective potential for $SU(3)$ for different values of the 
temperature.}
\label{fig:1}
\end{figure}

Immediately above the critical temperature, the $SU(3)$ potential 
presents a barrier between the old and the new vacua. This 
barrier, however, is very small and quickly disappears with the 
increasing of the temperature. One should notice that above 
$2T_c$ the changes in the potential are negligible.

  
\section{Langevin Evolution}

Let us now consider the real-time evolution of the order parameter for the
breakdown of $Z(N)$. We assume the system to be characterized
by a coarse-grained free energy
\begin{equation}
F(\phi,T)=\int d^3 x\left[\frac{B}{2} \, (\nabla \phi)^2+
V_{eff}(\phi,T)\right] \; ,
\end{equation}
where $V_{eff}(\phi,T)$ is the effective potential obtained in the last section, 
and $B = \pi^2 T/g^2$ for $SU(2)$ and $ B = 4 T/g^2$ for $SU(3)$. 
The time evolution of the order parameter and its approach to equilibrium will 
be dictated by the following Langevin equation
%
%
\begin{equation}
B  \left(\frac{\partial^2\psi}{\partial t^2} - \nabla^2 \psi 
\right) + \Gamma \, \frac{\partial\psi}{\partial t} + V'_{eff}(\psi) = 
\xi \; ,
\label{eq.1}
\end{equation}
where $g$ is the QCD coupling constant, and $\Gamma$ is the dissipation 
coefficient, which is usually taken to be a function of temperature only, 
$\Gamma=\Gamma (T)$. The function $\xi$ is a stochastic noise assumed 
to be gaussian and white so that 
$\langle\xi (\vec x, t)\rangle=0$ and 
$\langle \xi (\vec x, t)\xi(\vec x' ,t')\rangle=
2\Gamma \delta (\vec x- \vec x' )\delta (t - t')$. 
The noise and dissipation terms 
are originated from thermal and quantum fluctuations resulting 
either from self-interactions of the Polyakov loop field or 
from the coupling to diferent fields (such as chiral fields). The case 
with only first-order time derivative was considered in 
Ref.~\cite{Krein:2005wh}.

This description is admittedly very simplified. A more complete analysis 
should consider different contributions of noise and dissipation terms and 
memory kernels instead of simple Markovian terms proportional 
to the first time derivate of the field \cite{gleiser,rischke}. 
In general, one obtains a complicated dissipation kernel that 
simplifies to a multiplicative dissipation term which depends
quadratically on the amplitude of the field as 
$\Gamma_1(T)  \,\psi^2(\vec x,t) \, \dot L(\vec x,t)$ where 
$\Gamma_1$ is determined 
by imaginary terms of the effective action for $\psi$ and depends 
weakly (logaritmically) on the couplings. The fluctuation-dissipation 
theorem implies, then, that the noise term will 
also contain a multiplicative contribution of the form 
$\psi(\vec x,t)\xi(\vec x,t)$, and be in general non-Markovian. 
The white noise limit is reobtained only for very 
high temperatures. 

For the $SU(2)$ case we have fixed $\Gamma$ in the following way. 
We have used pure-gauge Euclidean lattice Monte Carlo simulations in the 
line discussed in Ref.~\cite{ogilvie}. In this approach, spinodal decomposition 
is obtained on the lattice performing local heat-bath updates of 
gauge field configurations at $\beta = 4/g^2 =3$, after thermalizing 
the lattice at $\beta = 4/g^2 = 2$. The critical value of $\beta$ for
deconfinement is found to be $\beta_d \sim 2.3$. $\Gamma$ is then 
extracted by comparing the short-time exponential growth of 
the correlation function $\langle L(k,t) L(-k,t)\rangle$ predicted by the 
lattice simulations \cite{AKT} and the Langevin description, assuming of 
course that both dynamics are the same. Making this comparison for the 
lowest lattice momentum mode, it is found that 
$\Gamma  = 7.6 \times 10^3 \, T^3 / \mu$,
where $\mu$ is a time scale relating Monte Carlo time and real time. Assuming
that typical thermalization times are of the order of a few fm/c, we obtain
$\Gamma \sim 10^3$~fm$^{-2}$.  

\vspace{0.55cm}
\begin{figure}[htbp]
\includegraphics[width=8cm]{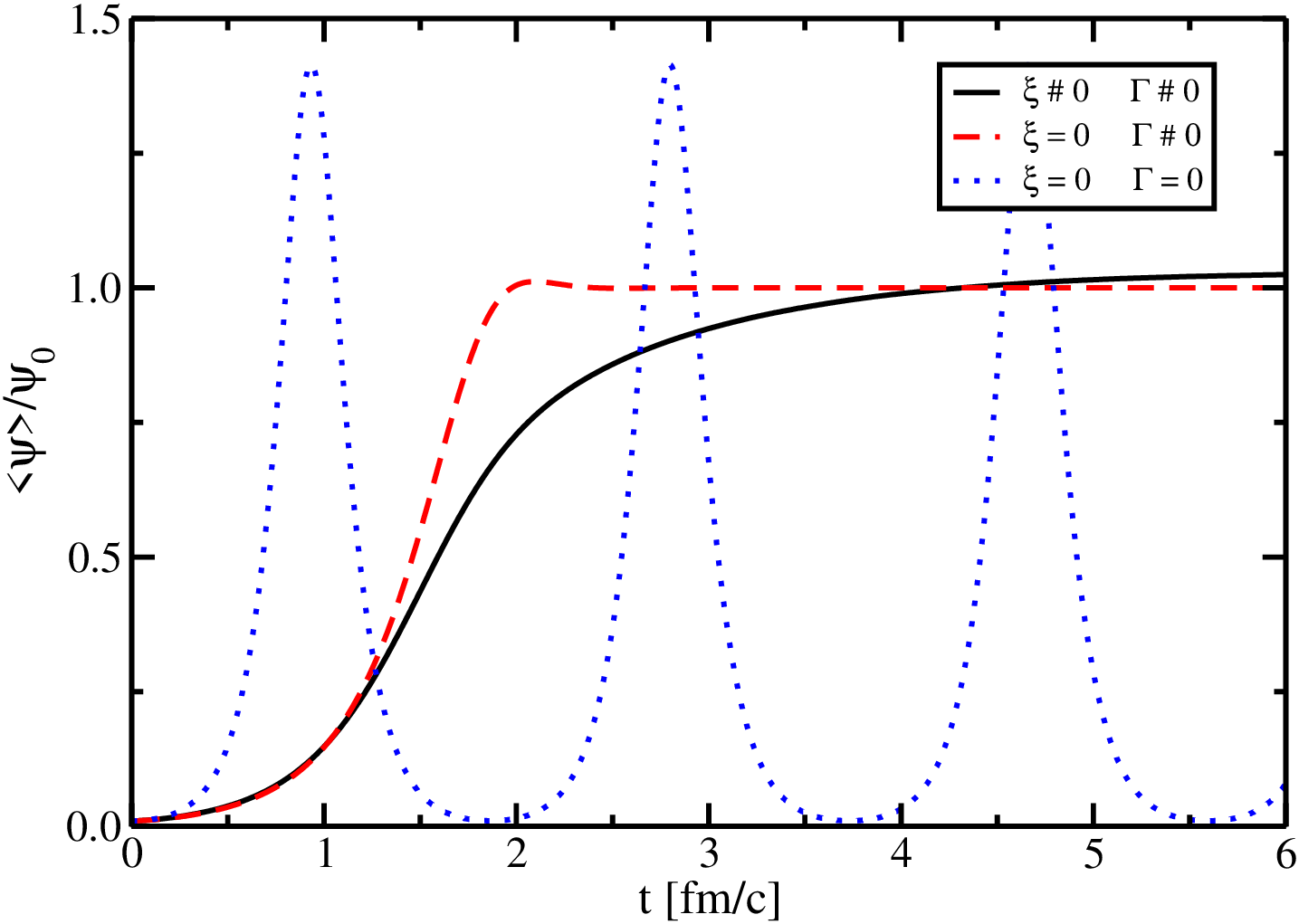}
\vspace{0.10cm}
\caption{Volume average of the $SU(2)$ order parameter normalized to the
$\psi_0 > 0$  minimum of the bare effective potential. }
\label{fig:2}
\end{figure}

In our numerical calculations we solve Eq. (\ref{eq.1}) on a cubic 
spacelike lattice with $64^3$ sites under periodic boundary 
conditions. We use a semi-implicit finite-diference scheme for 
the time evolution and a finite-difference Fast Fourier Transform for
the spatial dependence \cite{MIMC}. For $SU(2)$, the critical temperature is 
$T_d=302~$MeV \cite{karsch} and we obtain $M=775~$MeV. We took the 
average of several realizations with random initial configurations around 
$\psi \sim 0$.  We consider the time dependence of the volume average
of $\psi$
\begin{equation}
\langle \psi \rangle = \frac{1}{N^3} \sum_{ijk} \psi_{ijk}(t) \; ,
\label{vol_av}
\end{equation}
where $N$ is the number of lattice sites in each spatial direction, 
and $i,j,k = 1, \cdots, N$ are the lattice sites. In Fig.~\ref{fig:2} 
we plot $\langle \psi \rangle/\psi_0$, where 
$\psi_0 > 0$ is the positive minimum of the bare effective potential, 
for three situations: no dissipation and no noise (dotted curve), 
no noise (dashed curve) and full solution (solid curve). When considering 
noise, we have added the appropriate counterterms to make the equilibrium 
solution independent of the lattice spacing~\cite{CT}. All curves are 
for $T = 6.6 \, T_d$.

Clearly seen in Fig.~\ref{fig:2} is the large effect of dissipation, 
which delays the rapid exponential growth of the order parameter due 
to spinodal decomposition. The retardation seen here for the deconfinement 
transition is substantially larger than the corresponding delay seen for 
the chiral condensate evolution in Ref.~\cite{Fraga:2004hp}. The effect 
of noise is also in the direction of delaying equilibration, 
as expected. Also expected, and clearly shown in Fig.~\ref{fig:2}, is 
the effect of noise in the equilibrium value of $\psi$ which is larger 
than $\psi_0$. 

\vspace{0.55cm}
\begin{figure}[htbp]
\includegraphics[width=8cm]{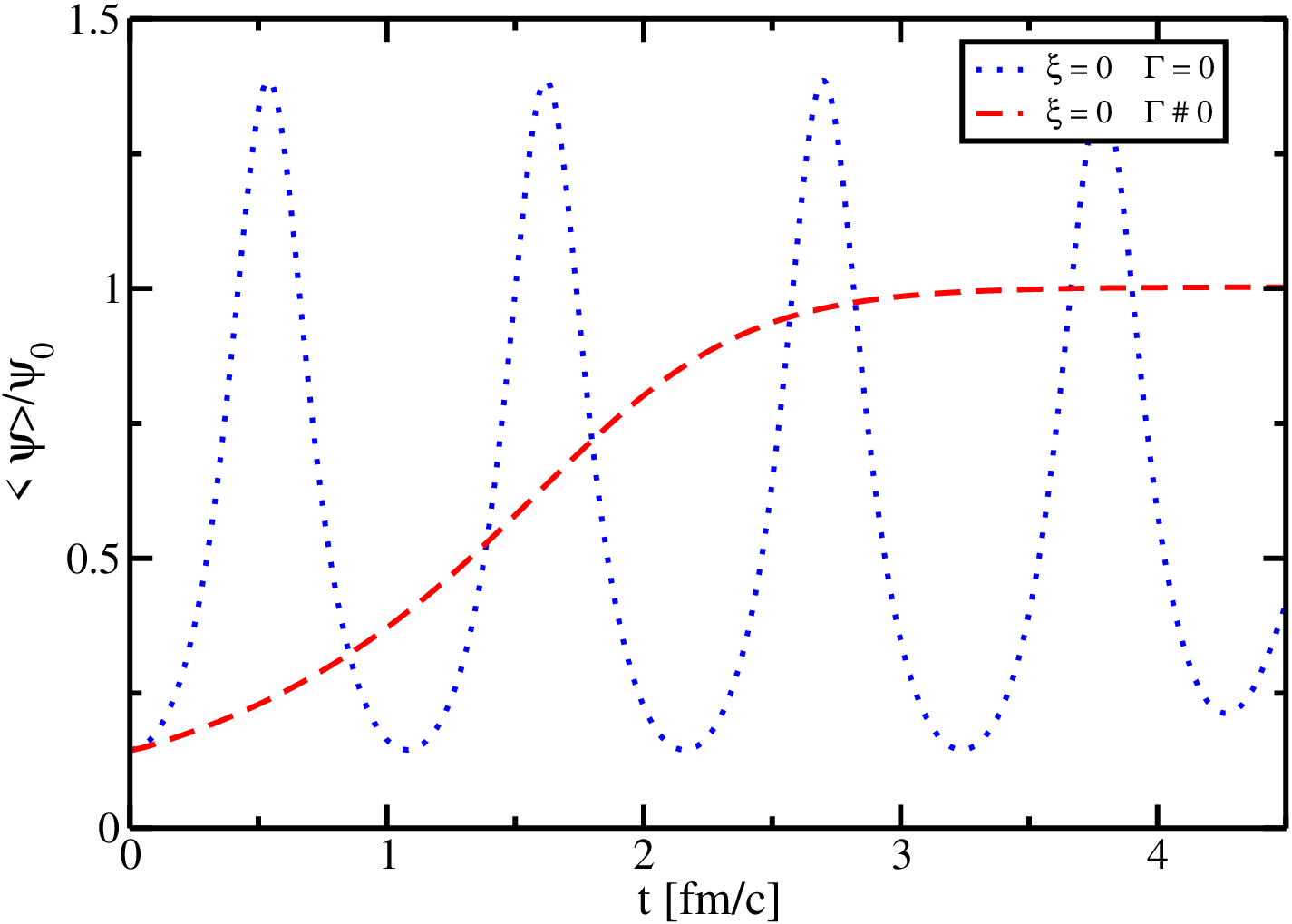}
\caption{Volume average of the $SU(3)$ order parameter normalized to the
$\psi_0 > 0$  minimum of the bare effective potential. }
\label{fig:3}
\end{figure}

For $SU(3)$ one has $T_d=263~$MeV \cite{Boyd:1996bx}, so that 
$M = 580~$MeV. 
The results of our simulations at $T= 6.6\,T_d$ are shown in 
Fig.~\ref{fig:3}. Here we are using the same lattice and same 
dissipation $\Gamma$ as for $SU(2)$. As seen in this figure, 
the effect of dissipation is even more dramatic than for 
$SU(2)$, with the proviso of course that we are using the value 
of $\Gamma$ extracted from $SU(2)$ lattice simulations. As mentioned 
earlier, immediately above the critical temperature the $SU(3)$ 
potential presents a barrier between a local minimum and an absolute 
minimum. However, this barrier has no effect on the delay seen in 
Fig.~\ref{fig:3}, since our simulations are done for high temperatures, 
$T \gg 2\,T_d$. We have not investigated the effect of noise in this 
case because the appropriate renormalization counter-terms for an 
effective potential with a first-order transition are not yet 
available~\cite{CT}. 


\section{Summary and outlook}

We have investigated the effects of dissipation and noise in the 
deconfinement transition of $SU(2)$ and $SU(3)$ pure gauge theories. 
We have used the effective model proposed in Ref.~\cite{ogilvie}, 
which combines phenomenological inputs with $Z(N)$ symmetry and some 
known features of the perturbative equation of state. The
model provides a reasonable representation of lattice results for 
the pure-gluon plasma equation of state in the temperature range 
between $T_c$ and $5T_c$ . We have performed numerical simulations 
for the evolution of the order parameter on a spatial cubic lattice 
using a local Langevin equation. We find that both dissipation and 
noise have dramatic effects on the spinodal decomposition of the 
$SU(2)$ order parameter, delaying considerably its thermalization. 
Dissipation effects are even larger for $SU(3)$.  

The present work must be improved in several aspects. Perhaps 
the most important one is in the method used to extract the dissipation 
coefficient $\Gamma$ \cite{AKT}. This was done using Euclidean 
lattice Monte Carlo simulations, in which spinodal decomposition 
of the order parameter is obtained performing local heat-bath updates 
of gauge field configurations above the deconfinement temperature. One 
of the major uncertainties in this approach is the relation between 
Monte Carlo updates and real time. Another source of uncertainties comes 
from a richer structure of noise and dissipation terms, including 
an evaluation of memory kernels. It is widely known that, in general, 
quantum corrections lead to complicated dissipation kernels that only 
in very special situations simplify to an additive noise term as used 
here. These issues will be considered in a future publication 
\cite{future}.

\acknowledgments
We thank G. Ananos for discussions and 
CAPES, CNPq, FAPERJ, FAPESP and FUJB/UFRJ for financial support.


\end{document}